\documentclass[aps,prl,twocolumn,letter,floatfix,superscriptaddress]{revtex4}
\pdfoutput=1
\usepackage{amssymb}
\usepackage{longtable}
\usepackage[toc,page]{appendix}
\usepackage{graphicx,color,float}
\usepackage{amsthm}
\usepackage{amsmath}
\usepackage[tight]{subfigure}
\usepackage[breaklinks=true,urlcolor=blue,bookmarks=true]{hyperref}
\hypersetup{pdfauthor={Jiang Qian},pdftitle={Quantum Signatures of the 
Optomechanical Instability}}

\begin{document}
\title{Quantum Signatures of the Optomechanical Instability}
\author{Jiang Qian}
\affiliation{Arnold Sommerfeld Center for Theoretical Physics, Center for 
NanoScience and Department of Physics, Ludwig-Maximilians-Universit\"at 
M\"unchen, Theresienstrasse 37, 80333, M\"unchen, Germany}

\author{A.~A. Clerk}
\affiliation{Department of Physics, McGill University, Montreal, Quebec, Canada H3A 2T8}

\author{K.~Hammerer}
\affiliation{Institute for Theoretical Physics, Institute for Gravitational Physics, Leibniz University Hanover, Callinstrasse 38, D-30167 Hanover, Germany}

\author{Florian Marquardt}
\affiliation{Institute for Theoretical Physics, Department of Physics, 
Friedrich-Alexander Universit\"at Erlangen-N\"urnberg, Staudtstr. 7, 91058 
Erlangen, Germany}
\affiliation{Max Planck Institute for the Science of Light, 
G\"unter-Scharowsky-Stra\ss e 1/Bau 24, 91058 Erlangen, Germany}

\begin{abstract}
In the past few years, coupling strengths between light and mechanical motion in
optomechanical setups have improved by orders of magnitude. Here we show that,
in the standard setup under continuous laser illumination, the steady state of the 
mechanical oscillator can develop a non-classical, strongly negative Wigner 
density if the optomechanical coupling is comparable to or larger than the 
optical decay rate and the mechanical frequency.  Because of its robustness, 
such a Wigner density can be mapped using optical homodyne tomography. This 
feature is observed near the onset of
the instability towards self-induced oscillations. We show that there are also
distinct signatures in the photon-photon correlation function $g^{(2)}(t)$ in
that regime, including oscillations decaying on a time scale not only much 
longer than the optical cavity decay time, but even longer than the mechanical decay time.
\end{abstract}

\date{\today}
\maketitle
By coupling optical and mechanical degrees of freedom, the emerging field of 
optomechanics provides exciting new opportunities to study the quantum 
mechanical behavior of macroscopic objects (for reviews see 
\cite{Marquardt2009,Kippenberg2008}).
Recent optomechanical cooling experiments are successfully bringing 
nanomechanical oscillators into their quantum mechanical ground state 
~\cite{TeufelCooling2011,PainterGround2011}.  The same optomechanical coupling 
also promises the possibility of single-quadrature measurements of the 
resulting mechanical quantum states with
the help of the light 
field~\cite{Braginsky,ClerkMarquardtQND,2010_HertzbergQND_NaturePhysics}.  For 
a reproducible and persistent quantum state, such measurements would result in 
an experimental determination of its full Wigner density via tomography, 
similar to what has been achieved in microscopic systems, for single ions or 
photons~\cite{IonTrapWigner,PhotonFockState}.  The recent advances in 
fabricating optomechanical devices have drastically improved coupling 
parameters, \emph{e.g.} for optomechanical crystals~\cite{PainterCrystal}, in 
microwave setups~\cite{TeufelCooling2011}, and other devices like GaAs 
disks~\cite{2010_Favero_GaAsDisk} or toroidal optical 
microcavity~\cite{Kippenberg2012}.  It will likely be possible relatively soon 
to achieve optomechanical coupling strengths $g_0$ at the single-photon level 
that are comparable to the optical cavity decay rate $\kappa$, a feat that has 
already been achieved in cold atom optomechanical 
systems\cite{Murch2008,EsslingerColdAtom2008}. This
regime of strongly nonlinear quantum optomechanics promises to pave the way 
towards generating and detecting novel quantum states in optomechanical 
systems. It is currently only beginning to be explored theoretically 
\cite{LudwigNJP2008,Rabl,NunnenkampPRL}, although very early work already 
discussed quantum optomechanical effects in the (unrealistic) absence of any 
dissipation~\cite{Mancini1997,Bose1997}.


In the classical regime, nonlinear dynamics is known to occur when
the system is driven by a blue detuned laser.  
When the input laser power crosses a certain threshold, the mechanical 
oscillator will undergo a Hopf bifurcation and start 
self-induced mechanical oscillations, a phenomenon termed ``parametric instability''
\cite{Braginsky1967,Kippenberg2005,Carmon2005,MarquardtPRL2006,Ludwig2008,GrudininPhononLaser2010}.  
The quantum dynamics of this regime has first been studied 
in~\cite{LudwigNJP2008}, and there is interesting synchronization behaviour for 
arrays of coupled oscillators of this type~\cite{OurPRL2011}.  



In this paper,  we show that, for strong optomechanical couplings $g_0$ 
 comparable to or greater than the optical decay rate $\kappa$ and mechanical 
frequency $\omega_M$~($g_0/\kappa\gtrsim 1,g_0^2/(\kappa\cdot\omega_M)\gtrsim 
1$), a large laser driving and an effectively zero temperature thermal bath, a 
non-classical
state of the mechanical oscillator with strongly negative Wigner 
density can be produced around
the onset of self-induced oscillations.  Because the state is time-independent, 
one may use single-quadrature homodyne tomography to experimentally reconstruct 
its non-classical Wigner density.

In addition, we propose to use the two-point photon correlation 
function $g^{(2)}(t)$ as an experimentally convenient probe for the peculiar
quantum dynamics near the bifurcation.  We identify two distinct signatures 
that enable experimentalists to reliably detect the onset and growth of the 
self-induced oscillation. We provide an explanation of the non-classical 
decay of $g^{(2)}(t)$ in both the red and blue-detuned regime.

\begin{figure*}
\begin{center}
\begin{tabular}{cccc}
\includegraphics[height=0.37\textwidth]{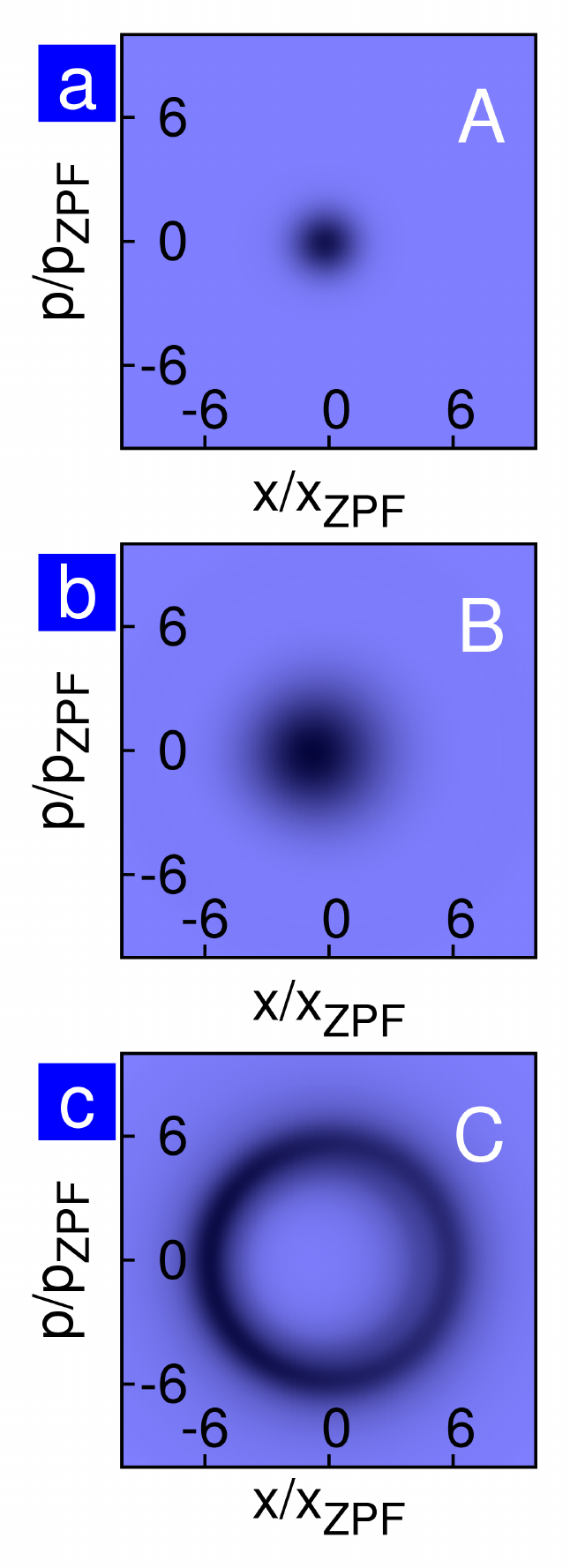} &
\includegraphics[height=0.37\textwidth]{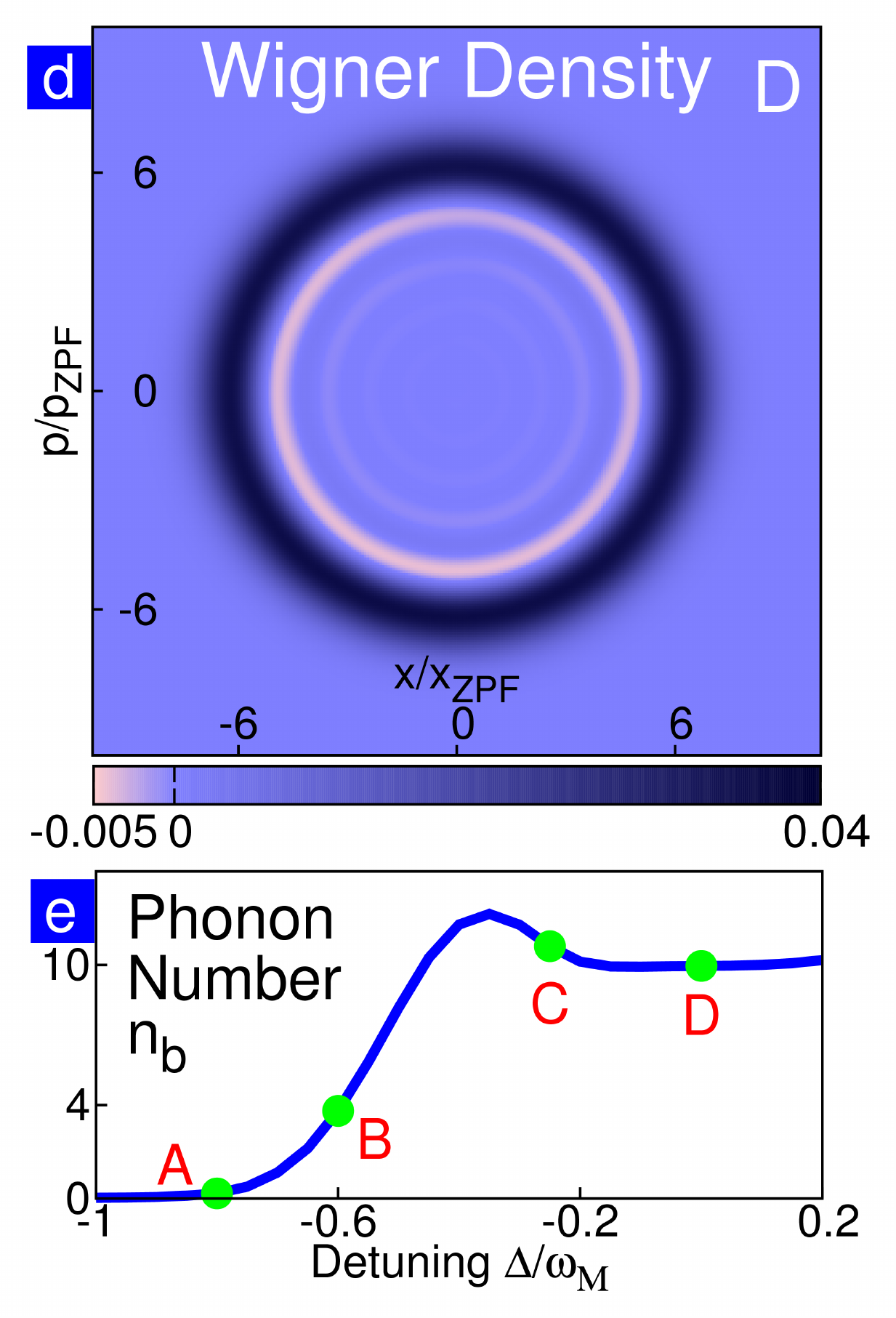} &
\includegraphics[height=0.37\textwidth]{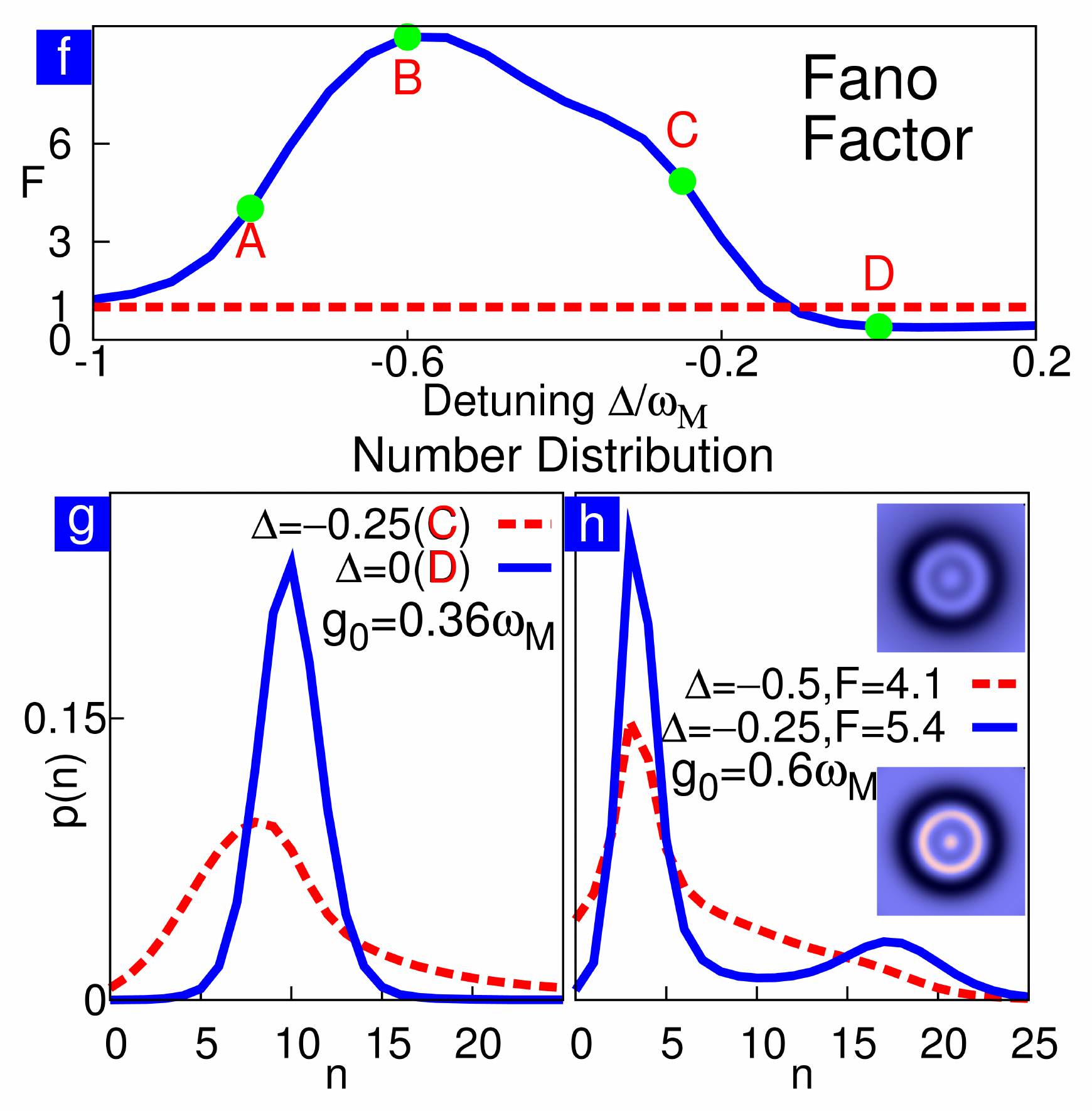} &
\includegraphics[height=0.37\textwidth]{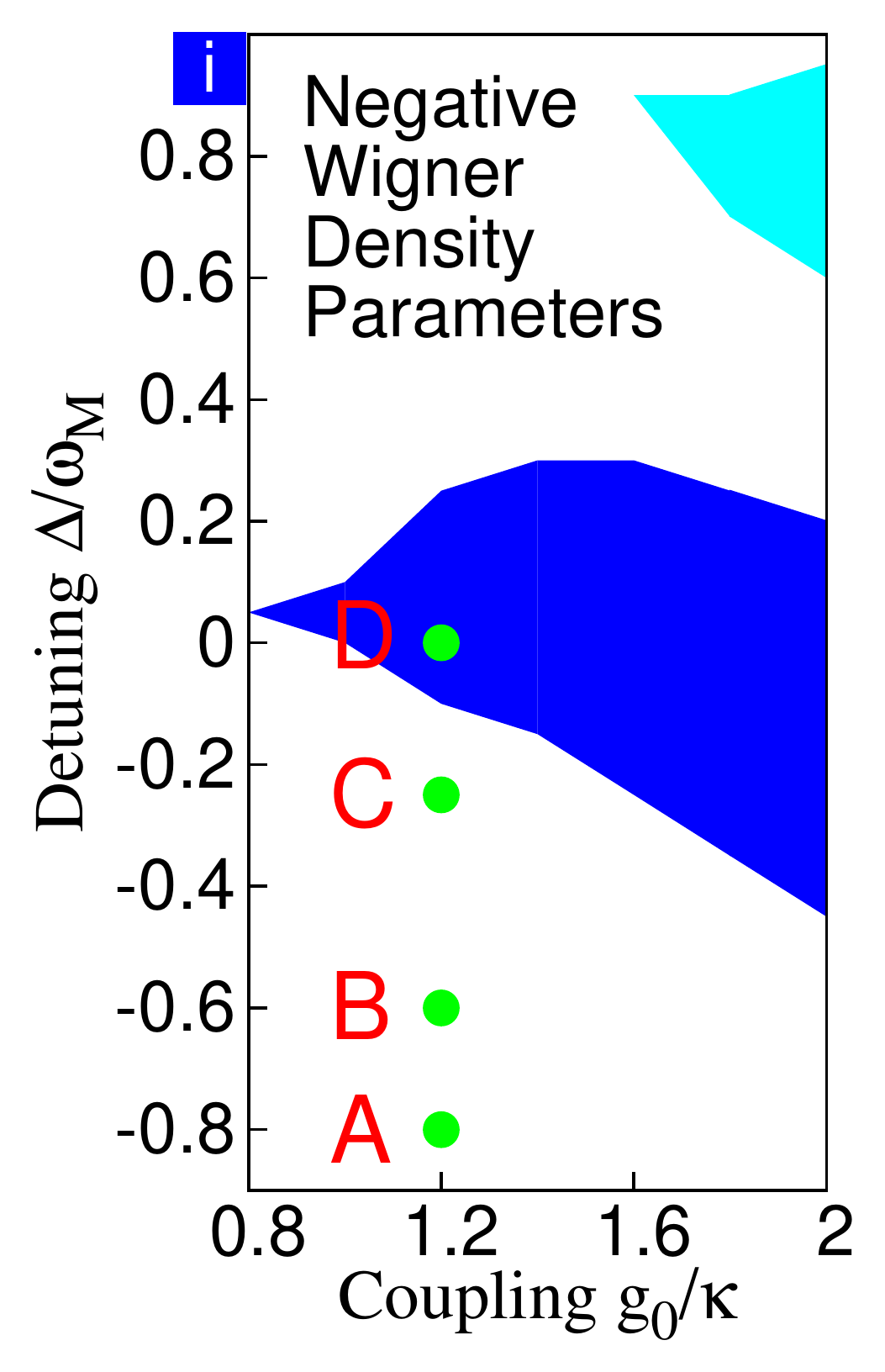}\\
\end{tabular}
\caption{\label{fig:2}Non-classical states in an optomechanical system.  The 
laser input $\alpha_L$ is held constant and the laser detuning $\Delta$ 
increases from the steady state ``A'' to ``D''. The mechanical Wigner densities 
of these states are shown in (a)-(d).  $x_{ZPF}$ and $p_{ZPF}$ are zero-point 
fluctuations of the oscillator's position and momentum, respectively. Plot (e) 
shows the start of the self-induced oscillation, where the phonon number $n_b$ 
of the oscillator rises quickly between state ``B'' and ``C''.  As the detuning 
further increases to ``D'', a non-classical mechanical quantum state with 
\emph{negative} mechanical Wigner density state appears, as shown in (d). In 
(f) the evolution of the mechanical Fano factor $F$ as a function of $\Delta$ 
is shown.  It dips below the Poisson value $1$ (dashed line) in non-classical 
state shown here.  In plot (g) and (h), we show that the negative Wigner 
density states have more sharply peaked phonon number distributions $p(n)$ 
compared with non-negative states. In (g) the $p(n)$ of state ``C'' and ``D''  
(plot (c),(d)) are compared. In (h), where $g_0=0.6\omega_M$, the negative 
state (solid line) has two clear peaks in $p(n)$, in contrast to a single 
smooth peak for the non-negative state (dashed line).  The Wigner density of 
these two states are shown as insets. Finally, in (i) we show two regions in 
the parameter space of detuning $\Delta$ and coupling $g_0$ where significant 
negative Wigner density states exist. States ``A''-``D'' are indicated here.  
In all plots other physical parameters are 
$g_0=0.36\omega_M,\kappa_M=0.3\omega_M,\Gamma_M=0.00147\omega_M,\alpha_L=0.311\omega_M$, 
except for (h), where $g_0=0.6\omega_M,\alpha_L=0.186\omega_M$. The 
intra-cavity photon number is $n_a\approx0.1$---$0.7$ when 
$g_0=0.36\omega_M,-\omega_M\le\Delta\le0$.}
\end{center}
\end{figure*}

Within the rotating wave approximation, an optomechanical system can be 
described by the following standard Hamiltonian:
\begin{equation}
\label{Ham}
\hat{H}=
\hbar(-\Delta+g_0(\hat{b}^{\dagger}+\hat{b}))\hat{a}^{\dagger}\hat{a}
+\hbar\omega_{M}\hat{b}^{\dagger}\hat{b}+\hbar\alpha_L(\hat{a}^{\dagger}
+\hat{a})+{\hat H}_{\rm diss}.
\end{equation}
Here ${\hat a}/{\hat b}$ are the operators for the photon/phonon modes, 
$\omega_M$ is the mechanical frequency and $\alpha_L$ is the laser driving 
amplitude.  $\Delta=\omega_{L}-\omega_{C}$ is the detuning of the laser from 
the cavity's \emph{unperturbed} resonance~(\emph{i.e.} evaluated for \emph{zero 
mechanical displacement}).  $g_0$ describes the strength of the optomechanical 
coupling at the single-photon level. 

When the dissipative terms in $H_{\rm diss}$ are taken into account, the 
density matrix $\hat{\rho}$ of the combined photon-phonon system evolves 
according to the quantum master equation:
\begin{equation}
\label{master}
\frac{d\hat{\rho}}{dt}=\mathcal{L}[\hat{\rho}]=
\frac{[\hat{H},\hat{\rho}]}{i\hbar}+
\Gamma\mathcal{D}[\hat{b},\hat{\rho}]+\kappa \mathcal{D}[\hat{a},\hat{\rho}].
\end{equation}
Here $\mathcal{L}$ is the quantum Liouville operator describing the time 
evolution of the density matrix $\hat{\rho}$, where we incorporate dissipation 
in the photon and phonon subsystems with decay rates $\kappa$ and $\Gamma$, 
respectively. The standard Lindblad term is given by
$\mathcal{D}[{\hat O},\hat{\rho}]=
\hat{O}\hat{\rho}\hat{O}^{\dagger}-
\frac{1}{2}(\hat{O}^{\dagger}\hat{O}\hat{\rho}
+\hat{\rho}\hat{O}^{\dagger}\hat{O})
$. Note that we will assume zero bath temperature in our simulations, 
which will be reachable to a good approximation when ${\rm GHz}$-frequency 
setups (\emph{e.g.} optomechanical crystals) are deployed in dilution 
refrigerator settings. In this paper, we are interested in the steady state 
solution of Eq.~\ref{master}, where all the transient dynamics has died out.  
This is obtained numerically by finding the density matrix satisfying 
$\mathcal{L}[\hat{\rho}]=0$ using the standard Arnoldi algorithm, as 
implemented in the ARPACK package.  Due to its persistence, this state is ideal 
for making homodyne measurements of its mechanical Wigner density, in contrast 
to transient scenarios.

Specifically we are interested in the mechanical Wigner density
$W_{\rm M}(x,p)=\frac{1}{\pi\hbar}\int_{-\infty}^{\infty}
\langle x-y|\hat{\rho}_{\rm M}|x+y\rangle e^{2ipy/\hbar}dy$, where 
$\hat{\rho}_{\rm M}$ is the mechanical density matrix, obtained by tracing out 
the optical degrees of freedom from $\hat{\rho}$. The Wigner density is the 
quantum analog of the classical Liouville phase space probability density. A 
negative Wigner density is a strong signature of a non-classical state. Early 
investigations~\cite{LudwigNJP2008} of the
optomechanical instability in the regime around $g_0\sim \kappa$ did not turned 
up nonclassical states.

In Fig~\ref{fig:2}, (a)-(e), we show the overall properties of the steady state 
solutions.  As we increase the laser detuning while keeping the input laser  
power fixed (points $\textrm{A}\to\textrm{B}\to\textrm{C}$), the phonon number 
in the mechanical oscillator rises sharply (plot (e)), signaling the onset of 
the self-induced oscillations. This is also reflected in the mechanical Wigner 
density $W_{\rm M}(x,p)$.  Below the onset (point ``A''), $W_{\rm M}(x,p)$ is a 
simple Gaussian, which starts to broaden just below the threshold, as the 
susceptibility of the system diverges and quantum fluctuations are strongly 
amplified (point ``B''). Above the threshold, we have a coherent state 
undergoing circular motion in phase space, but with an undetermined phase, 
which is the Wigner density observed at point ``C'' 
\cite{LudwigNJP2008,NunnenkampPRL}.

However, such a simple picture is inadequate for an optomechanical system with 
$g_0\sim\kappa$, \emph{i.e.}  when one approaches the optomechanical 
instability deep in the quantum regime~\footnote{Another necessary condition is 
that $g_0^2/(\kappa\cdot\omega_M)$ is not much smaller than one}.  In such a 
system, we observe that for a range of detuning $\Delta$ and laser driving 
$\alpha_L$, the mechanical self-induced oscillation produces strongly 
non-classical states with large negative areas in the Wigner density.  This can 
be seen in the example of Fig.~\ref{fig:2} (d).  Negative rims, shown in 
brighter color, develop at amplitudes slightly smaller than the average 
amplitude of oscillation.  Plots (f)-(h) in Fig.~\ref{fig:2} analyze negative 
states more deeply. In state ``D'', (f) shows the mechanical Fano factor 
$F=\frac{\langle\Delta n_b^2\rangle}{\langle n_b\rangle}$ dips below the 
coherent state value $1$, and its phonon number distribution (g) has a reduced 
variance. At larger coupling $g=0.6\omega_M$ (h), the negative state exhibits a 
sharp peak and a smoother one, as opposed to a single broader peak of the 
non-negative state~\footnote{Note here, due to the two-peak structure, the Fano 
factor of the negative Wigner density state remains above one}.  Overall, 
(f)-(h) show that the negative states are closer to a \emph{single} Fock state 
\emph{or} a superposition of \emph{few } Fock states as compared with a 
coherent state~\cite{Rodrigues2010}.  Note, however, the origin of this 
non-classical state is \emph{not }the same as that in the well-studied 
micromaser \cite{Filipowicz:86,Meystre:88,Krause87,Varcoe00}.  In the 
micromaser,  the mechanism relies crucially on the swapping of a single 
excitation between an excited atom and cavity over a fixed interaction time.  
These features are absent in our system.

Fig.~\ref{fig:2}~(i) maps out the regions in parameter space where negative 
Wigner densities occur. This `phase diagram' is shown as a function of the 
``quantum parameter'' $\zeta=\frac{g_0}{\kappa}$~\cite{LudwigNJP2008} and of 
the laser detuning $\Delta\omega_M$, at a fixed value of the laser driving 
strength $\alpha_L$. It has been obtained by solving for the steady state of 
the optomechanical system under constant illumination, and the Wigner density 
is considered as nonclassical if a sufficiently large area turns out to be 
negative. The threshold criterion is a negative area of at least 3\% of the 
positive area, and the minimum value being at least 5\% in absolute value of 
the maximum. The numerical results shown here indicate that, for the parameters 
considered here, starting at $\frac{g_0}{\kappa}=0.8$, the negative Wigner 
density states appear around detuning $\Delta/\omega_M=0$, and a second 
negative Wigner density region opens up at $\frac{g_0}{\kappa}=1.6$, around 
$\Delta/\omega_M=0.9$ at the first blue sideband, where the instability is 
driven efficiently. The phonon number distribution displays a pronounced 
narrowing, getting closer to a single or few mechanical Fock states. However, 
we find that still many photon/phonon levels are involved in the dynamics in 
the regime considered here, and there seems to be no simple explanation 
involving only a few levels. 

These \emph{steady-state} non-classical Wigner densities could be reconstructed 
via optomechanical Quantum Non-demolition quadrature 
detection~\cite{Braginsky,ClerkMarquardtQND} and subsequent quantum state 
tomography~\cite{2009_Lvovsky_Review_QuantumStateTomography}. This merely 
involves illumination with another amplitude-modulated laser beam for read-out, 
as explained in~\cite{ClerkMarquardtQND}.  When observed, these would provide 
an accessible example of non-classical states in a fabricated mesoscopic 
mechanical object.  To date, there has been no experimental observation of 
non-classical Wigner densities in the domain of micro- or nanomechanical 
structures. The experiment that came closest to
that goal, and in the process did produce nonclassical mechanical Fock states, 
employed a complex multi-layered superconducting circuit with piezoelectric 
coupling to a superconducting qubit and ultrafast pulse 
sequences~\cite{ClelandPiezo2010}. Furthermore in their setup the resonator 
lifetime is too short to permit the reconstruction of the full Wigner density.  
By contrast, once optomechanical parameters can be improved to reach the 
single-photon strong coupling regime, the scheme discussed here would be 
relatively straightforward, being based  on continuous laser illumination of an 
optomechanical setup whose fabrication is much less complex as it involves only 
one material. Recently a coupling $g_0/\kappa\approx0.007$ has been achieved in 
an optomechanical crystal system~\cite{painter2012} and further improvement is 
expected in that setup.  In addition, there is the possibility that the 
parameters required here may be reached in
cold atom optomechanical setups~\cite{Murch2008,EsslingerColdAtom2008}.

The full mechanical state reconstruction in the nonlinear quantum regime is an 
enticing and challenging goal. Nevertheless, it requires many experimental 
runs.  It will be helpful to have other means of optically probing the quantum 
dynamics of the system around the onset of the
instability. A very suitable probe for
the dynamics is provided by the two-point photon correlation function:
\begin{equation}
\label{equ:correlation}
g^{(2)}(t)=\frac
{\langle 
\hat{a}^{\dagger}_{\tau}\hat{a}^{\dagger}_{\tau+t}\hat{a}_{\tau+t}\hat{a}_{\tau}\rangle}
{(\langle \hat{a}^{\dagger}_{\tau}\hat{a}_{\tau} \rangle)^2}.
\end{equation}
$\langle\cdots\rangle$ denotes the average over $\hat{\rho}$. Here we employ 
the two-point correlator for the intra-cavity photon field, extractable from 
our numerical simulations. However, we emphasize that it can be shown using 
input-output theory~(See Appendix A-1) that Eq.~\ref{equ:correlation} also 
directly provides the $g^{(2)}$ function for the fluctuations of the output 
optical field.

In steady state, $g^{(2)}$ does not depend
on the initial time $\tau$. Photon correlations are readily accessible in 
quantum optics experiments today with single-photon detectors (\emph{e.g.}  
using a Hanbury-Brown Twiss setup), and they have been successfully employed to 
characterize the change of photonic statistics upon transmission through 
nonlinear systems. The most important example is photon anti-bunching in the 
resonance fluorescence of single photon emitters, which has recently also been 
predicted to occur in optomechanical systems for sufficiently strong 
coupling~\cite{Rabl}.

\begin{figure}
\centering
\includegraphics[width=0.45\textwidth]{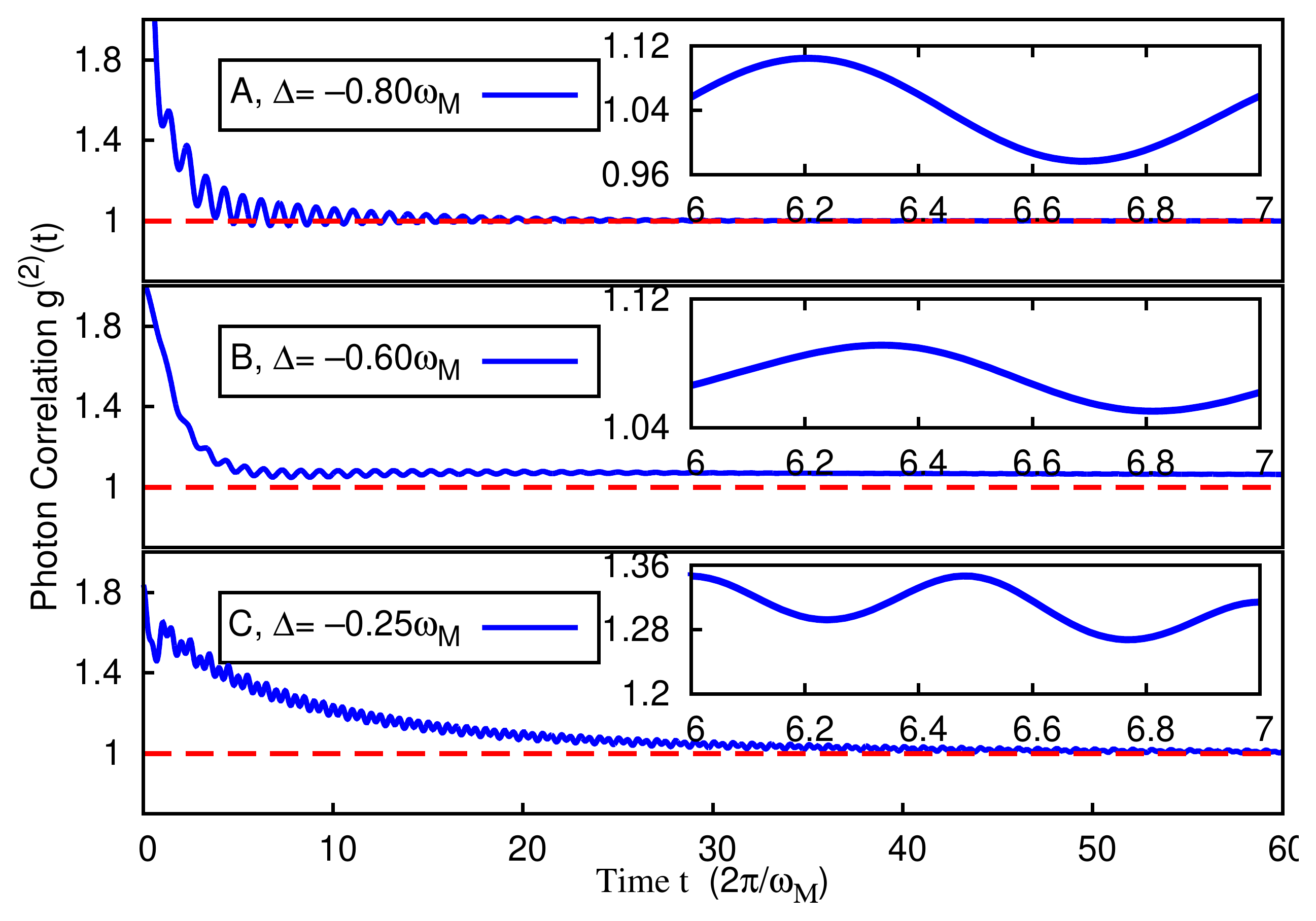}
\caption{\label{fig:4}Time-dependence of photon-photon correlations 
near the regime of quantum optomechanical oscillations.  ``A,B,C'' labels the 
same states as in Fig.~\ref{fig:2}.  
These plots show that there is a remarkably slow long-term decay near the onset 
of self-induced oscillations at point ``B'' (see main text).  Inset also shows 
the appearance of higher harmonics at point ``C''.}
\end{figure}

\begin{figure}
\centering
\includegraphics[width=0.45\textwidth]{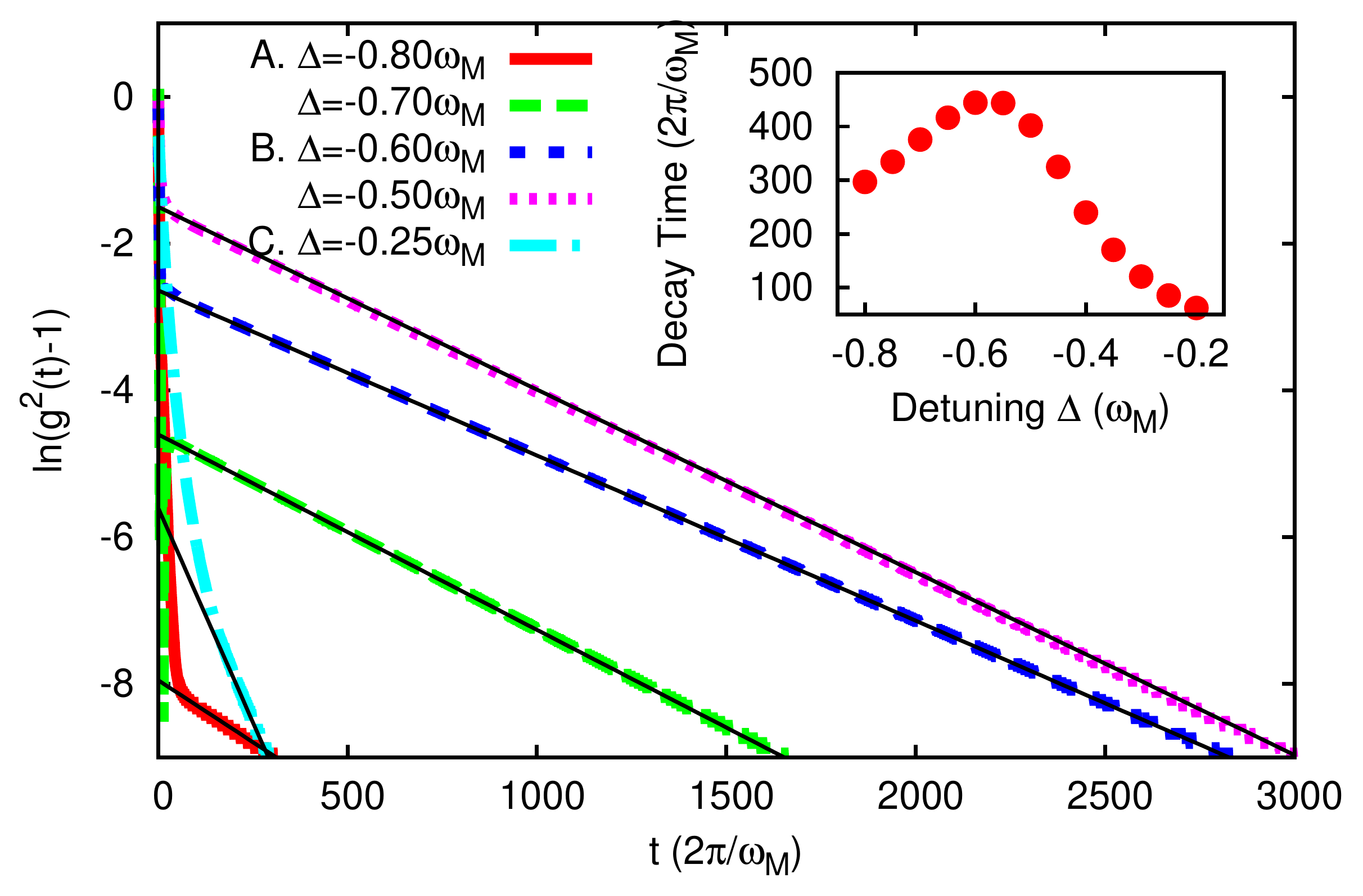}
\caption{\label{fig:new}Quantifying the slow approach of $g^{(2)}(t)\to1$ 
near the onset of the self-induced oscillations, as observed
in Fig.~\ref{fig:4}.
$g^{(2)}(t)-1$ obeys an exponential decay $e^{-t/\tau_g}$ in the long-time limit $t\to\infty$. The inset shows the decay time $\tau_g$ peaking toward very large values around point $B$, \emph{i.e.} $\Delta=0.6 \omega_M$.}
\end{figure}

As can be seen in Fig.~\ref{fig:4}, there are clear signatures in the photon 
correlator around the onset of parametric instability (point B).  In 
particular, $g^{(2)}(t)$ persists at some value above unity over a very long 
time~(middle panel, Fig.~\ref{fig:4}).
It can be proven~(see Appendix A-2) that as long as the steady state of the 
system is not degenerate, we always have 
$g^{(2)}(t)\to1+\alpha\exp{(-t/\tau_g)}$ in the long-time limit $t\to\infty$.  
Here the decay rate is $1/\tau_g=\mathrm{Re}{(\lambda_1)}$, where $\lambda_1$ 
is the eigenvalue of the Liouville operator $\mathcal{L}$ in Eq.~\ref{master} 
with the largest non-zero real part, characterizing the slowest decay in the 
system.  This can be verified by plotting $\ln(g^{(2)}(t)-1)$ to extract 
$\tau_g$, which indeed agrees with the $\lambda_1$ obtained from $\mathcal{L}$ 
(see Fig.~\ref{fig:new}). As can be seen in the inset, $\tau_g$ rises strongly 
around the start of the self-induced oscillation (point B). This is connected 
to
the fact that the overall mechanical damping rate goes to zero near 
the Hopf bifurcation~\cite{MarquardtPRL2006}.  

The second signature in $g^{(2)}$ is the appearance of higher harmonics when 
the self-induced oscillations are fully developed~(see insets of 
Fig.~\ref{fig:4}).  To understand these in a semiclassical picture, we 
approximate the photon correlator via the classical intensity correlator, 
${\langle|\alpha(t+\tau)|^2|\alpha(\tau)|^2\rangle_{\tau}}$. The light 
amplitude $\alpha(t)=e^{i\phi(t)}\sum\limits_{n}\alpha_n e^{in\omega_M t}$ is 
modulated harmonically by the mechanical oscillations, as detailed 
in~\cite{MarquardtPRL2006}. In Appendix A-3 we show that a fully developed 
mechanical self-induced oscillation results in higher harmonics in $g^{(2)}$.  
To understand the {\em decay} of the resulting oscillations in the $g^{(2)}$, 
we take into account the mechanical phase diffusion induced by the radiation 
pressure shot noise. \cite{VahalaPhaseNoise} presented the first analysis of 
the quantum contribution to phase diffusion in the parametric instability 
regime.  Here we follow a slightly modified approach.  The phase fluctuates 
according to $\delta{\phi}(t)=(m\omega_M A)^{-1} \int_{0}^{t}~dt'~\delta 
F(t')~\cos(\omega_M t')$, which yields:

\begin{displaymath}
\label{equ:phase_fluctuation}
\mathrm{Var}(\delta\phi(t))=\frac{1}{(m\omega_M 
A)^2}\frac{t}{4}\left(S_{FF}(\omega_M)+S_{FF}(-\omega_M)\right),\nonumber
\end{displaymath}

where $S_{FF}$ is the force noise spectrum (see~\cite{MarquardtPRL2007}).  Thus:

\begin{displaymath}
{\left\langle|\alpha(t+\tau)|^2|\alpha(\tau)|^2\right\rangle_{\tau}} = \sum_{n=-\infty}^{+\infty}Z_n e^{i n\omega_M t} e^{-n^2 \left\langle \delta\phi(t)^2 \right\rangle/2},
\end{displaymath}
where $Z_n=|\sum_{m=-\infty}^{\infty}\alpha_m\alpha^*_{m-n}|^2$. This theory 
explains qualitatively the shape of the correlator even deep in the quantum 
regime~(Appendix A-4).  Finally, we note that in the red detuned regime, the 
photon correlator decay can be described by the optomechanical cooling 
rate~(see Appendix A-4).

To summarize, in this paper we investigated quantum signatures of light and 
mechanics for an optomechanical system in the parametric instability regime.   
We found that, at strong optomechanical coupling 
($g_0\sim\kappa,g_0^2\sim(\kappa\cdot\omega_M)$), for a range of detuning and 
input power, the steady state mechanical Wigner density contains strong 
negative parts, signaling stable non-classical states.  Single-quadrature 
homodyne measurements can be used to reconstruct the Wigner density.  In 
addition, the  two-point photon correlator $g^{(2)}(t)$ displays two clear 
signatures near the onset of parametric instability. Finally we explained the 
slow long-time decay of the photon correlations as due to the mechanical phase 
diffusion induced by photon shot noise. One should note that experimental 
observation of some of these photon correlation features does
not require being in the nonlinear quantum regime and could succeed even in 
existing setups.

F.M. acknowledges the DFG (Emmy-Noether) and an ERC Starting Grant. F.M. and 
A.C. acknowledge the DARPA ORCHID program. J.Q. acknowledges the support of DFG 
SFB 631 and NIM. K.H acknowledges the support through QUEST.

\appendix
\appendixpage
\setcounter{equation}{0}
\renewcommand{\theequation}{A-\arabic{equation}}
\setcounter{figure}{0}
\renewcommand{\thefigure}{A-\arabic{figure}}
\renewcommand{\thesection}{A-\arabic{section}}
\section{A-1. Correlation Function: from Intra-cavity to Output 
Field}\label{appendix:correlation}
Here we summarize how the calculation of the $g^{(2)}$-function for the output 
field $\hat{a}_\mathrm{out}$ can be traced back to calculating $g^{(2)}$ for 
the intra-cavity field $\hat{a}$, following~\cite{Gardiner}. We are interested 
in the normally ordered two-time correlation function of the output field
\begin{equation}
g^{(2)}(t)=\frac{\langle 
\hat{a}^\dagger_\mathrm{out}(\tau)\hat{a}^\dagger_\mathrm{out}(t+\tau)\hat{a}_\mathrm{out}(t+\tau)\hat{a}_\mathrm{out}(\tau)\rangle}
{\langle 
\hat{a}^\dagger_\mathrm{out}(\tau)\hat{a}_\mathrm{out}(\tau)\rangle^2}.
\end{equation}
We substitute the input-output relation 
$\hat{a}_\mathrm{out}(t)=\hat{a}_\mathrm{in}(t)+\sqrt{\kappa}\hat{a}(t)$, and 
use that $\hat{a}^\dagger(\tau)$ commutes with 
$\hat{a}^\dagger_\mathrm{in}(t+\tau)$, and $\hat{a}(t+\tau)$ commutes with 
$\hat{a}_\mathrm{in}(\tau)$ for $t\geq 0$ as a consequence of causality, see 
\cite{Gardiner} for details. This permits us to bring the two time correlation 
function to a form where the $\hat{a}_\mathrm{in}^\dagger$ stand to the left, 
and the $\hat{a}_\mathrm{in}$ to the right of all other operators.  Moreover, 
note that for vacuum input 
$\hat{a}_\mathrm{in}\rho_\mathrm{in}=\rho_\mathrm{in}\hat{a}^\dagger_\mathrm{in}=0$.  
This ultimately establishes the identity
\begin{eqnarray}
&\langle 
\hat{a}^\dagger_\mathrm{out}(\tau)\hat{a}^\dagger_\mathrm{out}(t+\tau)\hat{a}_\mathrm{out}(t+\tau)\hat{a}_\mathrm{out}(\tau)\rangle
\nonumber\\
&=\kappa^2
\langle 
\hat{a}^\dagger(\tau)\hat{a}^\dagger(t+\tau)\hat{a}(t+\tau)\hat{a}(\tau)\rangle,
\end{eqnarray}
such that the normalized correlation function for the output field is 
\emph{identical} to the normalized correlation function of the intra-cavity 
field. This is what we calculated in Eq.~(3) of the main text.

\section{A-2. Proof Concerning the Longtime Limit of 
$g^{(2)}(t)$}\label{appendix:proof}
In this section we give a proof that the $g^{(2)}(t)$ defined in Eq.~3 of the 
main text, approaches one as $t\to\infty$.

We can rewrite the unnormalized correlation function (the numerator of Eq.~3 of
the main text) as follows:
\begin{eqnarray}
\label{equ:derivation}
g_{0}^{(2)}(t)&=&\mathrm{tr}[\hat{\rho}~\hat{a}^{\dagger}_{\tau}\hat{a}^{\dagger}_{t+\tau}\hat{a}_{t+\tau}\hat{a}_{\tau}]\nonumber\\
&=&\mathrm{tr}[(\hat{a}_{\tau}\hat{\rho}\hat{a}^{\dagger}_{\tau})~\hat{a}^{\dagger}_{t+\tau}\hat{a}_{t+\tau}]\nonumber\\
&=&\mathrm{tr}[(\hat{a}_{\tau}\hat{\rho}\hat{a}^{\dagger}_{\tau})~e^{\frac{i\hat{H}t}{\hbar}}~\hat{a}^{\dagger}_{\tau}\hat{a}_{\tau}~e^{-\frac{i\hat{H}t}{\hbar}}]\nonumber\\
&=&\mathrm{tr}[\hat{a}^{\dagger}_{\tau}\hat{a}_{\tau}~e^{\frac{-i\hat{H}t}{\hbar}}(\hat{a}_{\tau}\hat{\rho}\hat{a}^{\dagger}_{\tau})e^{\frac{i\hat{H}t}{\hbar}}]\nonumber\\
&=&\mathrm{tr}[\hat{a}^{\dagger}_{\tau}\hat{a}_{\tau}~e^{\mathcal{L}t}\hat{\rho}'].
\end{eqnarray}
Here $\hat{\rho}'=\hat{a}_{\tau}\hat{\rho}\hat{a}^{\dagger}_{\tau}$ and 
$e^{\mathcal{L}t}\hat{\rho}'$ is its time evolution under the quantum Liouville 
operator Eq.~2 of the main
text. In the last step we use the quantum regression approximation.

Let us now consider the right eigenvectors of $\mathcal{L}$:
\begin{equation}\label{equ:eigen}
\mathcal{L}\hat{\rho}_n=\lambda_n\hat{\rho}_n.
\end{equation}
Here we rank $\hat{\rho}_n$ in descending order of $\rm{Re}~\lambda_n$.  
Assuming the steady state $\lambda_0=0$ is not degenerate, we have 
$\rm{Re}~\lambda_n<0$ for $n>1$.  Since the trace is conserved in the time 
evolution according to Eq.~2 of the main text, $\rm{tr}(\hat{\rho}_n)=0$ for 
$n>0$ and $\rm{tr}(\hat{\rho}_0)=1$~(by normalization).

Expand the $\hat{\rho}'=\hat{a}_{\tau}\hat{\rho}\hat{a}^{\dagger}_{\tau}$ in 
eigenvectors $\hat{\rho}_n$:
\begin{eqnarray}
\hat{\rho}'&=&\sum_n c_n\hat{\rho}_n,\label{equ:rho-prime}\\
e^{\mathcal{L}t}\hat{\rho}'&=&\sum_n 
c_n\hat{\rho}_n~e^{\lambda_n}\label{equ:exponent}.
\end{eqnarray}
we can then evaluate the correlator in Eq.~\ref{equ:derivation} as 
$t\to\infty$:
\begin{eqnarray}
g_0^{(2)}(t\to\infty)&=&\lim_{t\to\infty} 
\mathrm{tr}[\hat{a}^{\dagger}_{\tau}\hat{a}_{\tau}~e^{\mathcal{L}t}\hat{\rho}']\nonumber\\
&=&\mathrm{tr}[\hat{a}^{\dagger}_{\tau}\hat{a}_{\tau}~c_0\hat{\rho}_0]\nonumber\\
&=&c_0~\mathrm{tr}[\hat{a}^{\dagger}_{\tau}\hat{a}_{\tau}\hat{\rho}_0]\nonumber\\
&=&c_0~\mathrm{tr}[\hat{a}^{\dagger}_{\tau}\hat{a}_{\tau}\hat{\rho}].
\end{eqnarray}
In the last step we use the fact that at time $\tau$ the system is in a steady 
state where the photon number no longer changes with time. Taking the trace of 
both sides of Eq.~\ref{equ:rho-prime} and utilizing the properties of 
$\rm{tr}(\hat{\rho}_n)$ discussed above, we have:
\begin{eqnarray}
c_0=\mathrm{tr}[\rho']=\mathrm{tr}[\hat{a}_{\tau}\hat{\rho}\hat{a}^{\dagger}_{\tau}]=\mathrm{tr}[\hat{a}^{\dagger}_{\tau}\hat{a}_{\tau}\hat{\rho}]=\langle 
\hat{a}^{\dagger}_{\tau}\hat{a}_{\tau} \rangle.
\end{eqnarray}
Thus we arrive at:
\begin{eqnarray}
g^{(2)}(t\to\infty)&=&\frac{g_0^{(2)}(t\to\infty)}{(\langle 
\hat{a}^{\dagger}_{\tau}\hat{a}_{\tau} \rangle)^2}\nonumber\\
&=&\frac{(\mathrm{tr}[\hat{a}^{\dagger}_{\tau}\hat{a}_{\tau}\hat{\rho}])^2}{(\langle 
\hat{a}^{\dagger}_{\tau}\hat{a}_{\tau} \rangle)^2}\nonumber\\
&=&1.
\end{eqnarray}
Finally, we point out that the leading term governing the asymptotic approach 
of $g^{(2)}\to1$ is $\lambda_1$ in Eq.~\ref{equ:eigen}, since it has the 
slowest exponential decay in Eq.~\ref{equ:exponent}. This gives the asymptotic 
behavior of $g^{(2)}(t)$ shown in the main text.

\section{A-3. Correlation Function: Semiclassical 
Picture}\label{appendix:semiclassical}
Under typical experimental conditions, when the classical self-induced 
oscillation starts, the mechanical motion is to a good approximation harmonic 
$x(t)\approx\bar{x}+A\cos(\omega_M t)$.  The laser amplitude, influenced by the 
mechanical oscillation, will contain higher harmonics~\cite{MarquardtPRL2006} 
$\alpha(t)=e^{i\phi(t)}\sum\limits_{n}\alpha_n e^{in\omega_M t}$, where
\begin{equation}
\label{equ:classical_solution}
\alpha_n=\frac{\alpha_L J_n(-g_0 A)}
{-n\omega_M+(\Delta+g_0\bar{x})+i\kappa/2}
\end{equation}
and $\phi(t)=g_0 A\sin(\omega_M t)$. Here we take the length unit to be the 
mechanical zero point fluctuation $x_{\textrm{ZPF}}$ and
frequency unit to be $\omega_M$. $J_n(x)$ is the n-th order Bessel function.  
 The oscillation amplitude $A$ and equilibrium position $\bar{x}$ can be 
determined self-consistently.
We can express $g^{(2)}(t)$ in terms of the coefficients in 
Eq.~\ref{equ:classical_solution}:
\begin{eqnarray}
\label{equ:z_classical}
Z_n=|\sum_{m=-\infty}^{\infty}\alpha_m\alpha^*_{m-n}|^2&,&
\langle|\alpha(\tau)|^2\rangle_{\tau}=\sum_{n=-\infty}^{\infty}
|\alpha_n|^2=Z_0.\nonumber\\
{\langle|\alpha(t+\tau)|^2|\alpha(\tau)|^2\rangle_{\tau}}&=&
Z_0+2\sum_{n=1}^{\infty}\cos(n\omega_M t)Z_n.
\end{eqnarray}
In this paper we're interested in the strongly quantum regime where $g_0\approx 
\kappa$, thus in the sideband resolved regime we have $g_0/\omega_M<1$.  From 
Eq.~\ref{equ:classical_solution} we see that only when $A\gg x_{\textrm{ZPF}}$ 
would there be significant contribution of higher harmonics in the light
amplitude $\alpha(t)$, which, as can be seen from Eq.~\ref{equ:z_classical} is 
also the condition of having higher harmonics in $g^{(2)}(t)$. This explains 
qualitatively the appearance of higher harmonics in the insets of Fig.~$2$ in
the main text when the quantum self-induced oscillation gains large amplitude.

\section{A-3. Correlation Function: Quantum 
Fluctuation}\label{appendix:fluctuation}
However, even when the self-induced oscillation has amplitude much larger than 
the $x_{\textrm{ZPF}}$, there are important quantum effects that are not 
accounted for by Eq.~\ref{equ:classical_solution} and 
Eq.~\ref{equ:z_classical}. As seen in Fig.~\ref{fig:5}, the classical 
solution~(bottom) is fully periodic, as there is a balance between the optical 
and mechanical dissipation and laser driving.  However, over the period of 60 
cycles, the amplitude of the quantum mechanical $g^{(2)}(t)$ decays 
significantly~(top three panels, blue curves). We can account for this decay by 
calculating the effect of shot noise fluctuations in the radiation pressure 
force $F(t)=(\hbar\omega_R/L)\hat{a}(t)^{\dagger}\hat{a}(t)$ on the phase 
$\phi$ of the mechanical oscillations:
\begin{figure}
\centering
\includegraphics[width=0.5\textwidth]{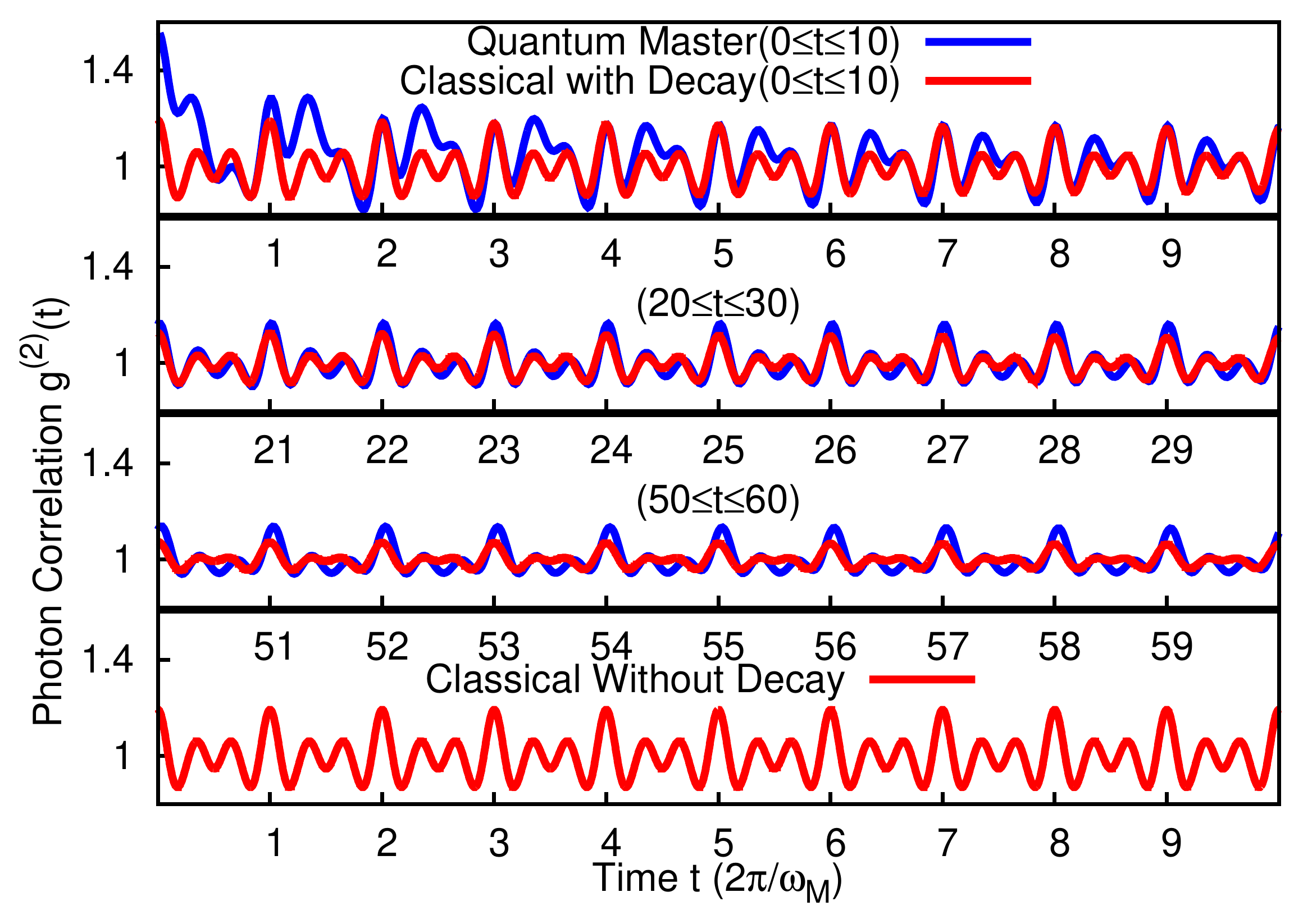}
\caption{\label{fig:5} Semiclassical approximation for photon correlations in 
the parametric instability regime of optomechanics. By adding the mechanical 
phase-diffusion to the classical light field dynamics (red), one can 
qualitatively account for the slow long-time decay of the photon correlator 
$g^{(2)}(t)$ in the full quantum simulation (blue). The lowest panel plots the 
classical solution without phase diffusion (here $\Delta=\omega_M$).
}
\end{figure}
\begin{figure}
\centering
\includegraphics[width=0.5\textwidth]{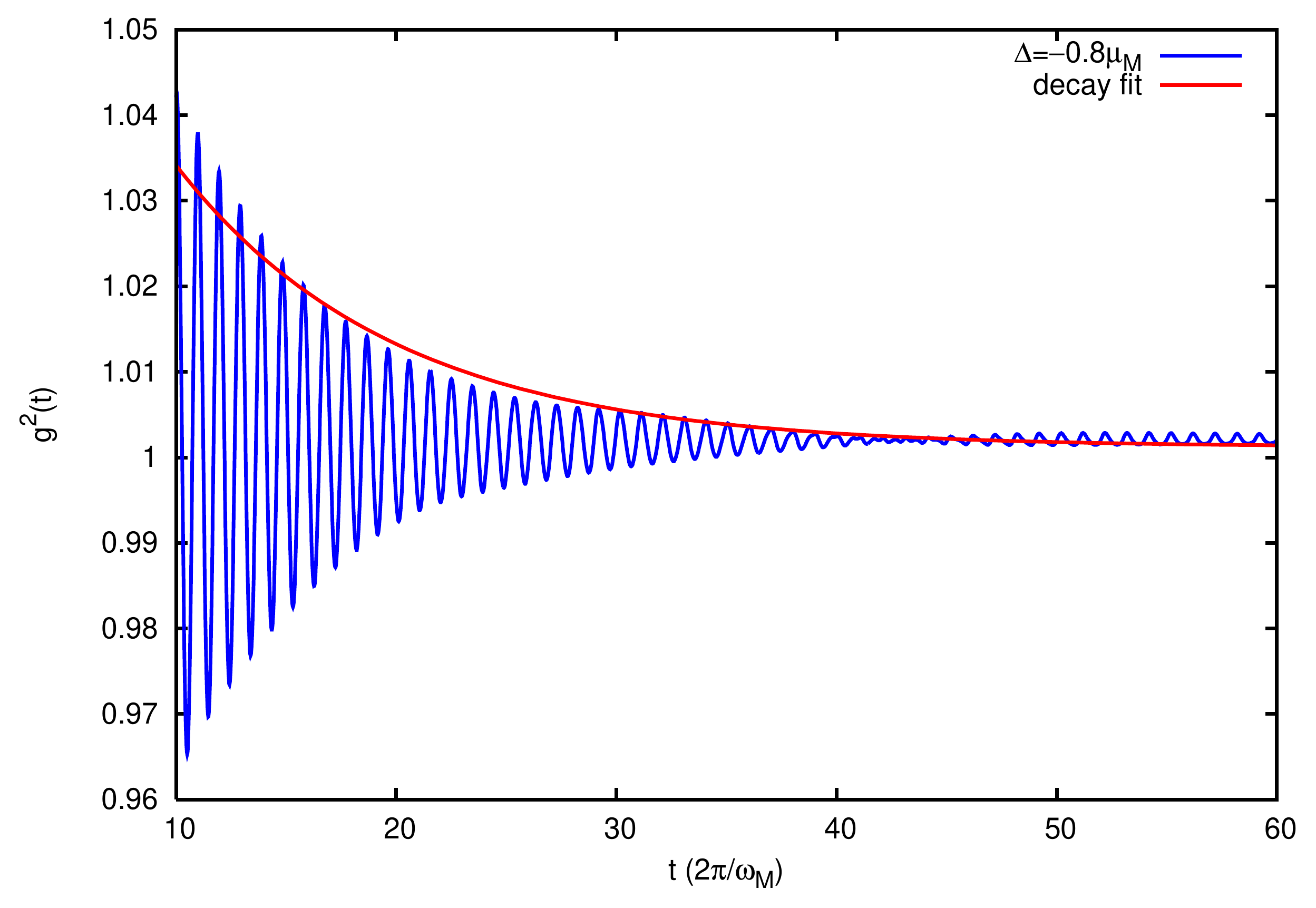}
\caption{\label{fig:6}The decay of the correlation function $g^{(2)}(t)$ in the 
red-detuned regime $\Delta=-0.8\omega_M$ can be qualitatively accounted for 
with $\Gamma_{opt}$ obtained from optomechanical cooling.}
\end{figure} 
\begin{eqnarray}
\delta{\phi}(t)&=&\frac{1}{m\omega_M A}\int_{0}^{t}~dt'~F(t')~\cos(\phi(t')),\\
\mathrm{Var}(\delta\phi(t))&=&
\frac{1}{(m\omega_M 
A)^2}\frac{t}{4}\left(S_{FF}(\omega_M)+S_{FF}(-\omega_M)\right).\nonumber
\end{eqnarray}
The noise spectrum of the radiation force is can be easily 
computed~\cite{MarquardtPRL2007}:
\begin{equation}
\label{equ:noise_spectrum}
S_{FF}(\omega)=
\left(\frac{\hbar\omega_R}{L}\right)\bar{n}_p\frac{\kappa}{(\omega+\Delta)^2+(\kappa/2)^2}.
\end{equation}
The fluctuations of the mechanical oscillator's phase feed back to the 
time-dependence of optical amplitude. Thus, under the semi-classical assumption 
where we take into account the photon shot noise but still treat the photon 
amplitude classically, we can modify Eq.~\ref{equ:z_classical} to be:
\begin{eqnarray}
{\langle|\alpha(t+\tau)|^2|\alpha(\tau)|^2\rangle_{\tau}}=
\sum_{n=-\infty}^{\infty} Z_n~e^{i n\omega_M t}\langle e^{i n 
(\delta\phi(t+\tau)-\delta\phi(t))}\rangle\nonumber&&\\
=Z_0+2\sum_{n=1}^{\infty}Z_n \cos(n\omega_M t) e^{\frac{-n^2}{2}
\mathrm{Var}(\delta\phi(t))}.&&\nonumber\\
\nonumber
\end{eqnarray}
Here we assume the phase fluctuation $\delta\phi$ is Gaussian. The result of 
this semi-classical accounting of the photon shot noise can be seen in the red 
curves in the top three panels of Fig.~\ref{fig:5}. We can see that this simple 
analysis can account qualitatively for the decay of the $g^{(2)}(t)$ in the 
large amplitude self-induced oscillation regime.

Finally, as we see in Fig.~\ref{fig:6}, there are also significant oscillation 
structure and decay for the $g^{(2)}(t)$ in the so-called red detuned regime, 
before the start of the self-induced oscillation. This regime cannot be 
understood at all by the classical picture, since classically the system has no
dynamics there. The oscillation in $g^{(2)}(t)$ can be understood as the 
dynamical response of the mechanical oscillator to the quantum fluctuation of 
the photon field. The decay can then be modeled using the theory of the 
optomechanical cooling of the mechanical oscillation in the red detuned regime, 
giving a cooling rate
\begin{equation}
\Gamma_{opt}=
\frac{x_{ZPF}^2}{\hbar^2}\left[S_{FF}(\omega_M)-S_{FF}(-\omega_M)\right].
\end{equation}
Here $S_{FF}$ is the same noise spectrum given by
Eq.~\ref{equ:noise_spectrum}.  As seen in Fig.~\ref{fig:6}, this rate gives a 
qualitative account for the decay rate of $g^{(2)}(t)$ in the red-detuned 
regime.
\bibliographystyle{apsrev}
\bibliography{wigner}

\begin{thebibliography}{10}%
\makeatletter
\providecommand \@ifxundefined [1]{%
 \ifx #1\undefined \expandafter \@firstoftwo
 \else \expandafter \@secondoftwo
\fi
}%
\providecommand \@ifnum [1]{%
 \ifnum #1\expandafter \@firstoftwo
 \else \expandafter \@secondoftwo
\fi
}%
\providecommand \enquote [1]{``#1''}%
\providecommand \bibnamefont  [1]{#1}%
\providecommand \bibfnamefont [1]{#1}%
\providecommand \citenamefont [1]{#1}%
\providecommand\href[0]{\@sanitize\@href}%
\providecommand\@href[1]{\endgroup\@@startlink{#1}\endgroup\@@href}%
\providecommand\@@href[1]{#1\@@endlink}%
\providecommand \@sanitize [0]{\begingroup\catcode`\&12\catcode`\#12\relax}%
\@ifxundefined \pdfoutput {\@firstoftwo}{%
 \@ifnum{\z@=\pdfoutput}{\@firstoftwo}{\@secondoftwo}%
}{%
 \providecommand\@@startlink[1]{\leavevmode\special{html:<a href="#1">}}%
 \providecommand\@@endlink[0]{\special{html:</a>}}%
}{%
 \providecommand\@@startlink[1]{%
  \leavevmode
  \pdfstartlink
   attr{/Border[0 0 1 ]/H/I/C[0 1 1]}%
   user{/Subtype/Link/A<</Type/Action/S/URI/URI(#1)>>}%
  \relax
 }%
 \providecommand\@@endlink[0]{\pdfendlink}%
}%
\providecommand \url  [0]{\begingroup\@sanitize \@url }%
\providecommand \@url [1]{\endgroup\@href {#1}{\urlprefix}}%
\providecommand \urlprefix [0]{URL }%
\providecommand \Eprint[0]{\href }%
\@ifxundefined \urlstyle {%
  \providecommand \doi [1]{doi:\discretionary{}{}{}#1}%
}{%
  \providecommand \doi [0]{doi:\discretionary{}{}{}\begingroup
  \urlstyle{rm}\Url }%
}%
\providecommand \doibase [0]{http://dx.doi.org/}%
\providecommand \Doi[1]{\href{\doibase#1}}%
\providecommand \bibAnnote [3]{%
  \BibitemShut{#1}%
  \begin{quotation}\noindent
    \textsc{Key:}\ #2\\\textsc{Annotation:}\ #3%
  \end{quotation}%
}%
\providecommand \bibAnnoteFile [2]{%
  \IfFileExists{#2}{\bibAnnote {#1} {#2} {\input{#2}}}{}%
}%
\providecommand \typeout [0]{\immediate \write \m@ne }%
\providecommand \selectlanguage [0]{\@gobble}%
\providecommand \bibinfo [0]{\@secondoftwo}%
\providecommand \bibfield [0]{\@secondoftwo}%
\providecommand \translation [1]{[#1]}%
\providecommand \BibitemOpen[0]{}%
\providecommand \bibitemStop [0]{}%
\providecommand \bibitemNoStop [0]{.\EOS\space}%
\providecommand \EOS [0]{\spacefactor3000\relax}%
\providecommand \BibitemShut [1]{\csname bibitem#1\endcsname}%
\bibitem{Marquardt2009}%
  \BibitemOpen
  \bibfield{author}{%
  \bibinfo {author} {\bibfnamefont{F.}~\bibnamefont{Marquardt}}\ and\ \bibinfo
  {author} {\bibfnamefont{S.~M.}\ \bibnamefont{Girvin}},\ }%
  \bibfield{journal}{%
  \bibinfo {journal} {Physics}\ }%
  \textbf{\bibinfo {volume} {2}},\ \bibinfo {pages} {40} (\bibinfo {year}
  {2009})%
  \bibAnnoteFile{NoStop}{Marquardt2009}%
\bibitem{Kippenberg2008}%
  \BibitemOpen
  \bibfield{author}{%
  \bibinfo {author} {\bibfnamefont{T.~J.}\ \bibnamefont{Kippenberg}}\ and\
  \bibinfo {author} {\bibfnamefont{K.~J.}\ \bibnamefont{Vahala}},\ }%
  \bibfield{journal}{%
  \bibinfo {journal} {Science}\ }%
  \textbf{\bibinfo {volume} {321}},\ \bibinfo {pages} {1172} (\bibinfo {year}
  {2008})%
  \bibAnnoteFile{NoStop}{Kippenberg2008}%
\bibitem{TeufelCooling2011}%
  \BibitemOpen
  \bibfield{author}{%
  \bibinfo {author} {\bibfnamefont{J.~D.}\ \bibnamefont{Teufel}}
  \emph{et~al.},\ }%
  \bibfield{journal}{%
  \bibinfo {journal} {Nature}\ }%
  \textbf{\bibinfo {volume} {475}},\ \bibinfo {pages} {359} (\bibinfo {year}
  {2011})%
  \bibAnnoteFile{NoStop}{TeufelCooling2011}%
\bibitem{PainterGround2011}%
  \BibitemOpen
  \bibfield{author}{%
  \bibinfo {author} {\bibfnamefont{J.}~\bibnamefont{Chan}}, \bibinfo {author}
  {\bibfnamefont{T.~P.~M.}\ \bibnamefont{Alegre}}, \bibinfo {author}
  {\bibfnamefont{A.~H.}\ \bibnamefont{Safavi-Naeini}}, \bibinfo {author}
  {\bibfnamefont{J.~T.}\ \bibnamefont{Hill}}, \bibinfo {author}
  {\bibfnamefont{A.}~\bibnamefont{Krause}}, \bibinfo {author}
  {\bibfnamefont{S.}~\bibnamefont{Groblacher}}, \bibinfo {author}
  {\bibfnamefont{M.}~\bibnamefont{Aspelmeyer}},\ and\ \bibinfo {author}
  {\bibfnamefont{O.}~\bibnamefont{Painter}},\ }%
  \bibfield{journal}{%
  \Doi{10.1038/nature10461}{\bibinfo {journal} {Nature}}\ }%
  \textbf{\bibinfo {volume} {478}},\ \bibinfo {pages} {89} (\bibinfo {year}
  {2011})%
  \bibAnnoteFile{NoStop}{PainterGround2011}%
\bibitem{Braginsky}%
  \BibitemOpen
  \bibfield{author}{%
  \bibinfo {author} {\bibfnamefont{V.~B.}\ \bibnamefont{Braginsky}}, \bibinfo
  {author} {\bibfnamefont{F.~Y.}\ \bibnamefont{Khalili}},\ and\ \bibinfo
  {author} {\bibfnamefont{K.~S.}\ \bibnamefont{Thorne}},\ }%
  \emph{\bibinfo {title} {Quantum Measurement}}\ (\bibinfo {publisher}
  {Cambridge University Press},\ \bibinfo {year} {1995})%
  \bibAnnoteFile{NoStop}{Braginsky}%
\bibitem{ClerkMarquardtQND}%
  \BibitemOpen
  \bibfield{author}{%
  \bibinfo {author} {\bibfnamefont{A.~A.}\ \bibnamefont{Clerk}}, \bibinfo
  {author} {\bibfnamefont{F.}~\bibnamefont{Marquardt}},\ and\ \bibinfo {author}
  {\bibfnamefont{K.}~\bibnamefont{Jacobs}},\ }%
  \bibfield{journal}{%
  \bibinfo {journal} {New Journal of Physics}\ }%
  \textbf{\bibinfo {volume} {10}},\ \bibinfo {pages} {095010} (\bibinfo {year}
  {2008})%
  \bibAnnoteFile{NoStop}{ClerkMarquardtQND}%
\bibitem{2010_HertzbergQND_NaturePhysics}%
  \BibitemOpen
  \bibfield{author}{%
  \bibinfo {author} {\bibfnamefont{J.~B.}\ \bibnamefont{Hertzberg}}, \bibinfo
  {author} {\bibfnamefont{T.}~\bibnamefont{Rocheleau}}, \bibinfo {author}
  {\bibfnamefont{T.}~\bibnamefont{Ndukum}}, \bibinfo {author}
  {\bibfnamefont{M.}~\bibnamefont{Savva}}, \bibinfo {author}
  {\bibfnamefont{A.~A.}\ \bibnamefont{Clerk}},\ and\ \bibinfo {author}
  {\bibfnamefont{K.~C.}\ \bibnamefont{Schwab}},\ }%
  \bibfield{journal}{%
  \bibinfo {journal} {Nature Physics}\ }%
  \textbf{\bibinfo {volume} {6}},\ \bibinfo {pages} {213} (\bibinfo {year}
  {2010})%
  \bibAnnoteFile{NoStop}{2010_HertzbergQND_NaturePhysics}%
\bibitem{IonTrapWigner}%
  \BibitemOpen
  \bibfield{author}{%
  \bibinfo {author} {\bibfnamefont{D.}~\bibnamefont{Leibfried}}, \bibinfo
  {author} {\bibfnamefont{D.~M.}\ \bibnamefont{Meekhof}}, \bibinfo {author}
  {\bibfnamefont{B.~E.}\ \bibnamefont{King}}, \bibinfo {author}
  {\bibfnamefont{C.}~\bibnamefont{Monroe}}, \bibinfo {author}
  {\bibfnamefont{W.~M.}\ \bibnamefont{Itano}},\ and\ \bibinfo {author}
  {\bibfnamefont{D.~J.}\ \bibnamefont{Wineland}},\ }%
  \bibfield{journal}{%
  \Doi{10.1103/PhysRevLett.77.4281}{\bibinfo {journal} {Phys. Rev. Lett.}}\ }%
  \textbf{\bibinfo {volume} {77}},\ \bibinfo {pages} {4281} (\bibinfo {year}
  {1996})%
  \bibAnnoteFile{NoStop}{IonTrapWigner}%
\bibitem{PhotonFockState}%
  \BibitemOpen
  \bibfield{author}{%
  \bibinfo {author} {\bibfnamefont{A.~I.}\ \bibnamefont{Lvovsky}}, \bibinfo
  {author} {\bibfnamefont{H.}~\bibnamefont{Hansen}}, \bibinfo {author}
  {\bibfnamefont{T.}~\bibnamefont{Aichele}}, \bibinfo {author}
  {\bibfnamefont{O.}~\bibnamefont{Benson}}, \bibinfo {author}
  {\bibfnamefont{J.}~\bibnamefont{Mlynek}},\ and\ \bibinfo {author}
  {\bibfnamefont{S.}~\bibnamefont{Schiller}},\ }%
  \bibfield{journal}{%
  \Doi{10.1103/PhysRevLett.87.050402}{\bibinfo {journal} {Phys. Rev. Lett.}}\
  }%
  \textbf{\bibinfo {volume} {87}},\ \bibinfo {pages} {050402} (\bibinfo {year}
  {2001})%
  \bibAnnoteFile{NoStop}{PhotonFockState}%
\bibitem{PainterCrystal}%
  \BibitemOpen
  \bibfield{author}{%
  \bibinfo {author} {\bibfnamefont{M.}~\bibnamefont{Eichenfield}}, \bibinfo
  {author} {\bibfnamefont{J.}~\bibnamefont{Chan}}, \bibinfo {author}
  {\bibfnamefont{R.~M.}\ \bibnamefont{Camacho}}, \bibinfo {author}
  {\bibfnamefont{K.~J.}\ \bibnamefont{Vahala}},\ and\ \bibinfo {author}
  {\bibfnamefont{O.}~\bibnamefont{Painter}},\ }%
  \bibfield{journal}{%
  \Doi{10.1038/nature08524}{\bibinfo {journal} {Nature}}\ }%
  \textbf{\bibinfo {volume} {462}},\ \bibinfo {pages} {78} (\bibinfo {year}
  {2009})%
  \bibAnnoteFile{NoStop}{PainterCrystal}%
\bibitem{2010_Favero_GaAsDisk}%
  \BibitemOpen
  \bibfield{author}{%
  \bibinfo {author} {\bibfnamefont{L.}~\bibnamefont{Ding}} \emph{et~al.},\ }%
  \bibfield{journal}{%
  \bibinfo {journal} {Phys. Rev. Lett.}\ }%
  \textbf{\bibinfo {volume} {105}},\ \bibinfo {pages} {263903} (\bibinfo {year}
  {2010})%
  \bibAnnoteFile{NoStop}{2010_Favero_GaAsDisk}%
\bibitem{Kippenberg2012}%
  \BibitemOpen
  \bibfield{author}{%
  \bibinfo {author} {\bibfnamefont{E.}~\bibnamefont{Verhagen}}, \bibinfo
  {author} {\bibfnamefont{S.}~\bibnamefont{Del\'eglise}}, \bibinfo {author}
  {\bibfnamefont{S.}~\bibnamefont{Weis}}, \bibinfo {author}
  {\bibfnamefont{A.}~\bibnamefont{Schliesser}},\ and\ \bibinfo {author}
  {\bibfnamefont{T.~K.}\ \bibnamefont{Kippenberg}},\ }%
  \bibfield{journal}{%
  \Doi{doi:10.1038/nature10787}{\bibinfo {journal} {Nature}}\ }%
  \textbf{\bibinfo {volume} {482}},\ \bibinfo {pages} {63} (\bibinfo {year}
  {2012})%
  \bibAnnoteFile{NoStop}{Kippenberg2012}%
\bibitem{Murch2008}%
  \BibitemOpen
  \bibfield{author}{%
  \bibinfo {author} {\bibfnamefont{K.~W.}\ \bibnamefont{Murch}}, \bibinfo
  {author} {\bibfnamefont{K.~L.}\ \bibnamefont{Moore}}, \bibinfo {author}
  {\bibfnamefont{S.}~\bibnamefont{Gupta}},\ and\ \bibinfo {author}
  {\bibfnamefont{D.~M.}\ \bibnamefont{Stamper-Kurn}},\ }%
  \bibfield{journal}{%
  \bibinfo {journal} {Nature Physics}\ }%
  \textbf{\bibinfo {volume} {4}},\ \bibinfo {pages} {561} (\bibinfo {year}
  {2008})%
  \bibAnnoteFile{NoStop}{Murch2008}%
\bibitem{EsslingerColdAtom2008}%
  \BibitemOpen
  \bibfield{author}{%
  \bibinfo {author} {\bibfnamefont{F.}~\bibnamefont{Brennecke}}, \bibinfo
  {author} {\bibfnamefont{S.}~\bibnamefont{Ritter}}, \bibinfo {author}
  {\bibfnamefont{T.}~\bibnamefont{Donner}},\ and\ \bibinfo {author}
  {\bibfnamefont{T.}~\bibnamefont{Esslinger}},\ }%
  \bibfield{journal}{%
  \Doi{10.1126/science.1163218}{\bibinfo {journal} {Science}}\ }%
  \textbf{\bibinfo {volume} {322}},\ \bibinfo {pages} {235} (\bibinfo {year}
  {2008})%
  \bibAnnoteFile{NoStop}{EsslingerColdAtom2008}%
\bibitem{LudwigNJP2008}%
  \BibitemOpen
  \bibfield{author}{%
  \bibinfo {author} {\bibfnamefont{M.}~\bibnamefont{Ludwig}}, \bibinfo {author}
  {\bibfnamefont{B.}~\bibnamefont{Kubala}},\ and\ \bibinfo {author}
  {\bibfnamefont{F.}~\bibnamefont{Marquardt}},\ }%
  \bibfield{journal}{%
  \Doi{10.1088/1367-2630/10/9/095013}{\bibinfo {journal} {New Journal of
  Physics}}\ }%
  \textbf{\bibinfo {volume} {10}},\ \bibinfo {pages} {095013} (\bibinfo {year}
  {2008})%
  \bibAnnoteFile{NoStop}{LudwigNJP2008}%
\bibitem{Rabl}%
  \BibitemOpen
  \bibfield{author}{%
  \bibinfo {author} {\bibfnamefont{P.}~\bibnamefont{Rabl}},\ }%
  \bibfield{journal}{%
  \Doi{10.1103/PhysRevLett.107.063601}{\bibinfo {journal} {Phys. Rev. Lett.}}\
  }%
  \textbf{\bibinfo {volume} {107}},\ \bibinfo {pages} {063601} (\bibinfo {year}
  {2011})%
  \bibAnnoteFile{NoStop}{Rabl}%
\bibitem{NunnenkampPRL}%
  \BibitemOpen
  \bibfield{author}{%
  \bibinfo {author} {\bibfnamefont{A.}~\bibnamefont{Nunnenkamp}}, \bibinfo
  {author} {\bibfnamefont{K.}~\bibnamefont{B\o{}rkje}},\ and\ \bibinfo {author}
  {\bibfnamefont{S.~M.}\ \bibnamefont{Girvin}},\ }%
  \bibfield{journal}{%
  \Doi{10.1103/PhysRevLett.107.063602}{\bibinfo {journal} {Phys. Rev. Lett.}}\
  }%
  \textbf{\bibinfo {volume} {107}},\ \bibinfo {pages} {063602} (\bibinfo {year}
  {2011})%
  \bibAnnoteFile{NoStop}{NunnenkampPRL}%
\bibitem{Mancini1997}%
  \BibitemOpen
  \bibfield{author}{%
  \bibinfo {author} {\bibfnamefont{S.}~\bibnamefont{Mancini}}, \bibinfo
  {author} {\bibfnamefont{V.~I.}\ \bibnamefont{Man'ko}},\ and\ \bibinfo
  {author} {\bibfnamefont{P.}~\bibnamefont{Tombesi}},\ }%
  \bibfield{journal}{%
  \bibinfo {journal} {Phys.\ Rev.\ A}\ }%
  \textbf{\bibinfo {volume} {55}},\ \bibinfo {pages} {3042} (\bibinfo {year}
  {1997})%
  \bibAnnoteFile{NoStop}{Mancini1997}%
\bibitem{Bose1997}%
  \BibitemOpen
  \bibfield{author}{%
  \bibinfo {author} {\bibfnamefont{S.}~\bibnamefont{Bose}}, \bibinfo {author}
  {\bibfnamefont{K.}~\bibnamefont{Jacobs}},\ and\ \bibinfo {author}
  {\bibfnamefont{P.~L.}\ \bibnamefont{Knight}},\ }%
  \bibfield{journal}{%
  \bibinfo {journal} {Phys.\ Rev.\ A}\ }%
  \textbf{\bibinfo {volume} {56}},\ \bibinfo {pages} {4175} (\bibinfo {year}
  {1997})%
  \bibAnnoteFile{NoStop}{Bose1997}%
\bibitem{Braginsky1967}%
  \BibitemOpen
  \bibfield{author}{%
  \bibinfo {author} {\bibfnamefont{V.}~\bibnamefont{Braginsky}}\ and\ \bibinfo
  {author} {\bibfnamefont{A.}~\bibnamefont{Manukin}},\ }%
  \bibfield{journal}{%
  \bibinfo {journal} {Soviet Physics JETP}\ }%
  \textbf{\bibinfo {volume} {25}},\ \bibinfo {pages} {653} (\bibinfo {year}
  {1967})%
  \bibAnnoteFile{NoStop}{Braginsky1967}%
\bibitem{Kippenberg2005}%
  \BibitemOpen
  \bibfield{author}{%
  \bibinfo {author} {\bibfnamefont{T.~J.}\ \bibnamefont{Kippenberg}}, \bibinfo
  {author} {\bibfnamefont{H.}~\bibnamefont{Rokhsari}}, \bibinfo {author}
  {\bibfnamefont{T.}~\bibnamefont{Carmon}}, \bibinfo {author}
  {\bibfnamefont{A.}~\bibnamefont{Scherer}},\ and\ \bibinfo {author}
  {\bibfnamefont{K.~J.}\ \bibnamefont{Vahala}},\ }%
  \bibfield{journal}{%
  \Doi{10.1103/PhysRevLett.95.033901}{\bibinfo {journal} {Phys. Rev. Lett.}}\
  }%
  \textbf{\bibinfo {volume} {95}},\ \bibinfo {pages} {033901} (\bibinfo {year}
  {2005})%
  \bibAnnoteFile{NoStop}{Kippenberg2005}%
\bibitem{Carmon2005}%
  \BibitemOpen
  \bibfield{author}{%
  \bibinfo {author} {\bibfnamefont{T.}~\bibnamefont{Carmon}}, \bibinfo {author}
  {\bibfnamefont{H.}~\bibnamefont{Rokhsari}}, \bibinfo {author}
  {\bibfnamefont{L.}~\bibnamefont{Yang}}, \bibinfo {author}
  {\bibfnamefont{T.}~\bibnamefont{Kippenberg}},\ and\ \bibinfo {author}
  {\bibfnamefont{K.}~\bibnamefont{Vahala}},\ }%
  \bibfield{journal}{%
  \bibinfo {journal} {Phys.\ Rev.\ Lett.}\ }%
  \textbf{\bibinfo {volume} {94}},\ \bibinfo {pages} {223902} (\bibinfo {year}
  {2005})%
  \bibAnnoteFile{NoStop}{Carmon2005}%
\bibitem{MarquardtPRL2006}%
  \BibitemOpen
  \bibfield{author}{%
  \bibinfo {author} {\bibfnamefont{F.}~\bibnamefont{Marquardt}}, \bibinfo
  {author} {\bibfnamefont{J.~G.~E.}\ \bibnamefont{Harris}},\ and\ \bibinfo
  {author} {\bibfnamefont{S.~M.}\ \bibnamefont{Girvin}},\ }%
  \bibfield{journal}{%
  \Doi{10.1103/PhysRevLett.96.103901}{\bibinfo {journal} {Phys. Rev. Lett.}}\
  }%
  \textbf{\bibinfo {volume} {96}},\ \bibinfo {pages} {103901} (\bibinfo {year}
  {2006})%
  \bibAnnoteFile{NoStop}{MarquardtPRL2006}%
\bibitem{Ludwig2008}%
  \BibitemOpen
  \bibfield{author}{%
  \bibinfo {author} {\bibfnamefont{C.}~\bibnamefont{Metzger}}, \bibinfo
  {author} {\bibfnamefont{M.}~\bibnamefont{Ludwig}}, \bibinfo {author}
  {\bibfnamefont{C.}~\bibnamefont{Neuenhahn}}, \bibinfo {author}
  {\bibfnamefont{A.}~\bibnamefont{Ortlieb}}, \bibinfo {author}
  {\bibfnamefont{I.}~\bibnamefont{Favero}}, \bibinfo {author}
  {\bibfnamefont{K.}~\bibnamefont{Karrai}},\ and\ \bibinfo {author}
  {\bibfnamefont{F.}~\bibnamefont{Marquardt}},\ }%
  \bibfield{journal}{%
  \bibinfo {journal} {Physical review Letters}\ }%
  \textbf{\bibinfo {volume} {101}},\ \bibinfo {pages} {133903} (\bibinfo {year}
  {2008})%
  \bibAnnoteFile{NoStop}{Ludwig2008}%
\bibitem{GrudininPhononLaser2010}%
  \BibitemOpen
  \bibfield{author}{%
  \bibinfo {author} {\bibfnamefont{I.~S.}\ \bibnamefont{Grudinin}}, \bibinfo
  {author} {\bibfnamefont{H.}~\bibnamefont{Lee}}, \bibinfo {author}
  {\bibfnamefont{O.}~\bibnamefont{Painter}},\ and\ \bibinfo {author}
  {\bibfnamefont{K.~J.}\ \bibnamefont{Vahala}},\ }%
  \bibfield{journal}{%
  \bibinfo {journal} {Physical Review Letters}\ }%
  \textbf{\bibinfo {volume} {104}},\ \bibinfo {pages} {083901} (\bibinfo {year}
  {2010})%
  \bibAnnoteFile{NoStop}{GrudininPhononLaser2010}%
\bibitem{OurPRL2011}%
  \BibitemOpen
  \bibfield{author}{%
  \bibinfo {author} {\bibfnamefont{G.}~\bibnamefont{Heinrich}}, \bibinfo
  {author} {\bibfnamefont{M.}~\bibnamefont{Ludwig}}, \bibinfo {author}
  {\bibfnamefont{J.}~\bibnamefont{Qian}}, \bibinfo {author}
  {\bibfnamefont{B.}~\bibnamefont{Kubala}},\ and\ \bibinfo {author}
  {\bibfnamefont{F.}~\bibnamefont{Marquardt}},\ }%
  \bibfield{journal}{%
  \Doi{10.1103/PhysRevLett.107.043603}{\bibinfo {journal} {Phys. Rev. Lett.}}\
  }%
  \textbf{\bibinfo {volume} {107}},\ \bibinfo {pages} {043603} (\bibinfo {year}
  {2011})%
  \bibAnnoteFile{NoStop}{OurPRL2011}%
\bibitem{Note1}%
  \BibitemOpen
  \bibinfo {note} {Another necessary condition is that $g_0^2/(\kappa \cdot
  \omega _M)$ is not much smaller than one}%
  \bibAnnoteFile{NoStop}{Note1}%
\bibitem{Note2}%
  \BibitemOpen
  \bibinfo {note} {Note here, due to the two-peak structure, the Fano factor of
  the negative Wigner density state remains above one}%
  \bibAnnoteFile{NoStop}{Note2}%
\bibitem{Rodrigues2010}%
  \BibitemOpen
  \bibfield{author}{%
  \bibinfo {author} {\bibfnamefont{D.~A.}\ \bibnamefont{Rodrigues}}\ and\
  \bibinfo {author} {\bibfnamefont{A.~D.}\ \bibnamefont{Armour}},\ }%
  \bibfield{journal}{%
  \Doi{10.1103/PhysRevLett.104.053601}{\bibinfo {journal} {Phys. Rev. Lett.}}\
  }%
  \textbf{\bibinfo {volume} {104}},\ \bibinfo {pages} {053601} (\bibinfo {year}
  {2010})%
  \bibAnnoteFile{NoStop}{Rodrigues2010}%
\bibitem{Filipowicz:86}%
  \BibitemOpen
  \bibfield{author}{%
  \bibinfo {author} {\bibfnamefont{P.}~\bibnamefont{Filipowicz}}, \bibinfo
  {author} {\bibfnamefont{J.}~\bibnamefont{Javanainen}},\ and\ \bibinfo
  {author} {\bibfnamefont{P.}~\bibnamefont{Meystre}},\ }%
  \bibfield{journal}{%
  \Doi{10.1364/JOSAB.3.000906}{\bibinfo {journal} {J. Opt. Soc. Am. B}}\ }%
  \textbf{\bibinfo {volume} {3}},\ \bibinfo {pages} {906} (\bibinfo {month}
  {Jun}\ \bibinfo {year} {1986})%
  \bibAnnoteFile{NoStop}{Filipowicz:86}%
\bibitem{Meystre:88}%
  \BibitemOpen
  \bibfield{author}{%
  \bibinfo {author} {\bibfnamefont{P.}~\bibnamefont{Meystre}}, \bibinfo
  {author} {\bibfnamefont{G.}~\bibnamefont{Rempe}},\ and\ \bibinfo {author}
  {\bibfnamefont{H.}~\bibnamefont{Walther}},\ }%
  \bibfield{journal}{%
  \Doi{10.1364/OL.13.001078}{\bibinfo {journal} {Opt. Lett.}}\ }%
  \textbf{\bibinfo {volume} {13}},\ \bibinfo {pages} {1078} (\bibinfo {month}
  {Dec}\ \bibinfo {year} {1988})%
  \bibAnnoteFile{NoStop}{Meystre:88}%
\bibitem{Krause87}%
  \BibitemOpen
  \bibfield{author}{%
  \bibinfo {author} {\bibfnamefont{J.}~\bibnamefont{Krause}}, \bibinfo {author}
  {\bibfnamefont{M.~O.}\ \bibnamefont{Scully}},\ and\ \bibinfo {author}
  {\bibfnamefont{H.}~\bibnamefont{Walther}},\ }%
  \bibfield{journal}{%
  \Doi{10.1103/PhysRevA.36.4547}{\bibinfo {journal} {Phys. Rev. A}}\ }%
  \textbf{\bibinfo {volume} {36}},\ \bibinfo {pages} {4547} (\bibinfo {month}
  {Nov}\ \bibinfo {year} {1987})%
  \bibAnnoteFile{NoStop}{Krause87}%
\bibitem{Varcoe00}%
  \BibitemOpen
  \bibfield{author}{%
  \bibinfo {author} {\bibfnamefont{B.~T.~H.}\ \bibnamefont{Varcoe}}, \bibinfo
  {author} {\bibfnamefont{S.}~\bibnamefont{Brattke}},\ and\ \bibinfo {author}
  {\bibfnamefont{H.}~\bibnamefont{Walther}},\ }%
  \bibfield{journal}{%
  \bibinfo {journal} {Journal of Optics B: Quantum and Semiclassical Optics}\
  }%
  \textbf{\bibinfo {volume} {2}},\ \bibinfo {pages} {154} (\bibinfo {year}
  {2000})%
  \bibAnnoteFile{NoStop}{Varcoe00}%
\bibitem{2009_Lvovsky_Review_QuantumStateTomography}%
  \BibitemOpen
  \bibfield{author}{%
  \bibinfo {author} {\bibfnamefont{A.~I.}\ \bibnamefont{Lvovsky}}\ and\
  \bibinfo {author} {\bibfnamefont{M.~G.}\ \bibnamefont{Raymer}},\ }%
  \bibfield{journal}{%
  \bibinfo {journal} {Reviews of Modern Physics}\ }%
  \textbf{\bibinfo {volume} {81}},\ \bibinfo {pages} {299} (\bibinfo {year}
  {2009})%
  \bibAnnoteFile{NoStop}{2009_Lvovsky_Review_QuantumStateTomography}%
\bibitem{ClelandPiezo2010}%
  \BibitemOpen
  \bibfield{author}{%
  \bibinfo {author} {\bibfnamefont{A.~D.}\ \bibnamefont{O'Connell}}, \bibinfo
  {author} {\bibfnamefont{M.}~\bibnamefont{Hofheinz}}, \bibinfo {author}
  {\bibfnamefont{M.}~\bibnamefont{Ansmann}}, \bibinfo {author}
  {\bibfnamefont{R.~C.}\ \bibnamefont{Bialczak}}, \bibinfo {author}
  {\bibfnamefont{M.}~\bibnamefont{Lenander}}, \bibinfo {author}
  {\bibfnamefont{E.}~\bibnamefont{Lucero}}, \bibinfo {author}
  {\bibfnamefont{M.}~\bibnamefont{Neeley}}, \bibinfo {author}
  {\bibfnamefont{D.}~\bibnamefont{Sank}}, \bibinfo {author}
  {\bibfnamefont{H.}~\bibnamefont{Wang}}, \bibinfo {author}
  {\bibfnamefont{M.}~\bibnamefont{Weides}}, \bibinfo {author}
  {\bibfnamefont{J.}~\bibnamefont{Wenner}}, \bibinfo {author}
  {\bibfnamefont{J.~M.}\ \bibnamefont{Martinis}},\ and\ \bibinfo {author}
  {\bibfnamefont{A.~N.}\ \bibnamefont{Cleland}},\ }%
  \bibfield{journal}{%
  \Doi{10.1038/nature08967}{\bibinfo {journal} {Nature}}\ }%
  \textbf{\bibinfo {volume} {464}},\ \bibinfo {pages} {697} (\bibinfo {year}
  {2010})%
  \bibAnnoteFile{NoStop}{ClelandPiezo2010}%
\bibitem{painter2012}%
  \BibitemOpen
  \bibfield{author}{%
  \bibinfo {author} {\bibfnamefont{J.}~\bibnamefont{Chan}}, \bibinfo {author}
  {\bibfnamefont{A.~H.}\ \bibnamefont{Safavi-Naeini}}, \bibinfo {author}
  {\bibfnamefont{J.~T.}\ \bibnamefont{Hill}}, \bibinfo {author}
  {\bibfnamefont{S.}~\bibnamefont{Meenehan}},\ and\ \bibinfo {author}
  {\bibfnamefont{O.}~\bibnamefont{Painter}},\ }%
  \bibfield{journal}{%
  \Doi{10.1063/1.4747726}{\bibinfo {journal} {Applied Physics Letters}}\ }%
  \textbf{\bibinfo {volume} {101}},\ \bibinfo {eid} {081115} (\bibinfo {year}
  {2012})%
  \bibAnnoteFile{NoStop}{painter2012}%
\bibitem{VahalaPhaseNoise}%
  \BibitemOpen
  \bibfield{author}{%
  \bibinfo {author} {\bibfnamefont{K.~J.}\ \bibnamefont{Vahala}},\ }%
  \bibfield{journal}{%
  \Doi{10.1103/PhysRevA.78.023832}{\bibinfo {journal} {Phys. Rev. A}}\ }%
  \textbf{\bibinfo {volume} {78}},\ \bibinfo {pages} {023832} (\bibinfo {year}
  {2008})%
  \bibAnnoteFile{NoStop}{VahalaPhaseNoise}%
\bibitem{MarquardtPRL2007}%
  \BibitemOpen
  \bibfield{author}{%
  \bibinfo {author} {\bibfnamefont{F.}~\bibnamefont{Marquardt}}, \bibinfo
  {author} {\bibfnamefont{J.~P.}\ \bibnamefont{Chen}}, \bibinfo {author}
  {\bibfnamefont{A.~A.}\ \bibnamefont{Clerk}},\ and\ \bibinfo {author}
  {\bibfnamefont{S.~M.}\ \bibnamefont{Girvin}},\ }%
  \bibfield{journal}{%
  \Doi{10.1103/PhysRevLett.99.093902}{\bibinfo {journal} {Phys. Rev. Lett.}}\
  }%
  \textbf{\bibinfo {volume} {99}},\ \bibinfo {pages} {093902} (\bibinfo {year}
  {2007})%
  \bibAnnoteFile{NoStop}{MarquardtPRL2007}%
\bibitem{Gardiner}%
  \BibitemOpen
  \bibfield{author}{%
  \bibinfo {author} {\bibfnamefont{C.~W.}\ \bibnamefont{Gardiner}}\ and\
  \bibinfo {author} {\bibfnamefont{P.}~\bibnamefont{Zoller}},\ }%
  \emph{\bibinfo {title} {Quantum Noise}},\ \bibinfo {edition} {3rd}\ ed.\
  (\bibinfo {publisher} {Springer},\ \bibinfo {year} {2010})%
  \bibAnnoteFile{NoStop}{Gardiner}%
\end{thebibliography}%

\end{document}